\documentstyle[12pt]{article}

\date{}
\begin{document}

\title{Geometric Quantization of free fields in space of motions\thanks{%
this work is supported by NSF of China, Pan Den Plan of China and LWTZ -1298
of Chinese Academy of Sciences}}
\author{{\ Ming-Xue Shao\thanks{{\protect\small E-mail: shaomx@itp.ac.cn}},
Zhong-Yuan Zhu\thanks{{\protect\small E-mail: zzy@itp.ac.cn}}} \\
{\small CCAST(World Laboratory), P.O.Box 8730, Beijing, 100080, P.R.China}\\
{\small Institute of Theoretical Physics, Academia Sinica, P.O.Box 2735,
Beijing 100080, P.R.China.}}
\maketitle

\begin{abstract}
Via K$\ddot{a}$hker polarization we geometrically quantize free fields in the 
spaces of motions,  namely the space of solutions of equations of motion. We 
obtain the correct results just as that given by the canonical quantization. 
Since we follow the method of covariant symplectic current proposed by Crnkovic, 
Witten and Zuckerman  et al, the canonical commutator  we obtained are  
naturally invariant under proper Lorentz transformation and the discrete parity 
and time transverse transformations, as well as the equations of motion. 
\end{abstract}
 
\section{Introduction}

The symplectic geometrical description of classical mechanics and geometric
quantization \cite{Wood Sniatycki}are essentially globalizations of, 
respectively, Hamiltonian
mechanics and canonical quantization. Geometric quantization is also
considered as a so far most mathematically thorough approach to
quantization. This formalism has been shown to provide an natural way to
investigate global and geometrical properties of physical systems with
geometrical invariance, such as Chern-Simons theory \cite{Witten-cs}, anyon
system \cite{YZL}, and so on. The method of geometry also provide principles by 
itself to clarify some ambiguities in traditional canonical quantization.
 
 It is  well known, not as the path integral which can keep the classical 
symmetries very well, the traditional descriptions of geometrical and canonical 
formalism of classical theories are not manifestly covariant because one has to 
explicitly single out a "time" coordinate to define the canonical conjugate 
momenta and the initial data of systems.  In the end of eighties 
E.Witten \cite{Witten} and
G.Zuckerman \cite{Zuckman} and C.Crnkovic \cite{Crnkovic1} et al. suggested a
manifestly covariant geometric descriptions, where  the space of solutions of 
the equation of
motion (called space of motion $M$) is taken as the state space\cite{souriau}. 
This definition is
independent of any special time choice so that is manifestly covariant. Then
this method were used to discuss Yang-Mills theory, general relativity, string 
theory and theory of supersymmetry. 
The presymplectic form constructed by covariant symplectic current  is not only 
invariant under Lorentz transformation, gauge transformation and diffeomorphism 
transformation but also
has zero component along gauge and diffeomorphism orbits.

In order to use the symplectic structure to study the quantization of fields 
theory, it is convenient to express the symplectic form in terms of ladder 
fields in momenta space by the Fourier decompositions of fields. This will be 
discussed in this paper in detail.  This step is necessary for K$\ddot{a}$hler 
polarization \cite{Wood Sniatycki} in geometric quantization. Via the covariant 
method we directly obtain the Poisson brackets of fields.  The symplectic form 
is invariant under proper Lorentz transformation and the discrete parity and 
time transverse transformations(LPT),
so naturally are the Poisson brackets and the corresponding quantum commutator.

This paper is organized as followings. In section 2 we review the Crnkovic, 
Zuckerman and Witten's  descriptions of  space of motion. In section 3,
we obtain the expressions of symplectic form in the solution space in terms of 
ladder fields in momentum space of free fields. Then we calculate the Poisson 
brackets of fields by the symplectic form. Finally we complete the geometric 
quantization in space of motion via  the K$\ddot{a}$hler polarizationand. 

%
\section{Crnkovic-Witten-Zuckerman's Covariant Symplectic Current Description}

In this section we will briefly review the description of covariant phase
space developed by E.Witten \cite{Witten}, G.Zuckerman \cite{Zuckman} and
C.Crnkovic \cite{Crnkovic1} et al. With $\phi ^a$ a collection of fields
which form a representation of its symmetry group of the theory the
variation of a local Lagrangian density $L=L(\phi ^a,\partial _\mu \phi ^a)$
is 
\begin{equation}
\delta L=\partial _\mu j^\mu +(E-L)_a\delta {\phi ^a}  , \label{001}
\end{equation}
where 
\begin{equation}
j^\mu =\frac{\partial L}{\partial \partial _\mu \phi ^a}\delta \phi ^a
,
\label{002}
\end{equation}
and 
\begin{equation}
(E-L)_a=\frac{\partial L}{\partial \phi ^a}-\partial {\mu }\frac{\partial L}{%
\partial \partial _\mu \phi ^a}
\end{equation}
the extreme of the action leads to the Euler-Lagrangian equations $(E-L)_a=0$%
. Now the phase space is defined to be the space $Z$ of solutions of the
Euler-Lagrange equations which is one to one correspondence to the
traditional phase space with a fixed time. Before introducing the symplectic
form on $Z$, we first define the vector field and the differential forms on $%
Z$. For simplicity, we assume field $\phi ^a$ to be a real scalar field $%
\phi $. For a fixed $x\in M$ with $M$ the base manifold on which the field $%
\phi $ is defined, a mapping from $Z$ to real numbers that assign to every
function $\phi $ its value at $x$ is a function on $Z$. a function ${\hat{x}}%
:Z\mapsto R$, where ${\hat{x}}(\phi )=\phi (x)$.

For every $\phi \in Z$, every small displacement $\delta \phi$ which is the
solution of the linearized equation of motion, is defined to be the vector
of the tangent space $T_{\phi}Z$ to $Z$ at $\phi$. The tangent vector field $%
\triangle$ is the section of tangent bundle $T_Z$.

1-forms, being elements of the dual vector space to the tangent vector
field, will map vector fields $\triangle $ into functions on $Z$. For vector
field $\triangle $ and $x\in M$, we define functions $\triangle _x$ on $Z$ 
\[
\triangle _x(\phi )\equiv \delta \phi (x) 
\]
where $\delta \phi (x)$ is a number, the value of the displacement $\delta
\phi $ at $x$.we define the 1-form $x^{*}$ on $Z$ by demanding the
contraction of $x^{*}$ with $\triangle $ gives function $\triangle _x$ 
$$
x^{*}(\triangle )=\triangle _x 
$$

The variation $\delta $ in (\ref{001}) can be interpreted as the infinite
dimensional exterior derivative operator and $\delta \phi $ the one-form on
phase space. So in the following for simplicity we just use $\delta \phi $
as both vector in $T_Z$ and 1-form in $T^{*}Z$ and $\delta $ as the exterior
derivative. There will not be confusion. The properties of differential
forms on $T^{*}Z$ is similar to that of the finite dimensional case.

Now the covariant symplectic current is defined to be 
\begin{equation}
\delta j^\mu =\delta \frac{\partial L}{\partial \partial _\mu \phi ^a}\wedge
\delta \phi ^a.  \label{X104}
\end{equation}
With $\Sigma $ to be the spacelike supersurface of space-time manifold, the
presymplectic form is defined to be 
\begin{equation}
\Omega =\int_{}^{}d\Sigma _\mu \delta j^\mu ,  \label{31presym}
\end{equation}
which is obviously  closed  in the covariant phase
space. To ensure the presymplectic form is well defined on solution space $Z$%
, it is needed to prove that $\Omega $ is independent of the choice of the
spacelike supersurface. Consider the presymplectic form defined on two
spacelike surface $\Sigma _1$ and $\Sigma _2$, we get 
\begin{equation}
\Omega (\Sigma _1)-\Omega (\Sigma _2)=\int_{}^{}d\Sigma _{1\mu }\delta j^\mu
-\int_{}^{}d\Sigma _{2\mu }\delta j^\mu =\int_VdV\partial _\mu \delta j^\mu
-\int_{\Sigma _3}d\Sigma _{3\mu }\delta j^\mu
\end{equation}
where $V$ is the four dimensional area enveloped by $\Sigma _1$ $\Sigma _2$,
and $\Sigma _3$, $\Sigma _3=\partial V-\Sigma _2-\Sigma _1$ is a time-like
surface at infinity. Using (\ref{001}), the above expression becomes 
\begin{equation}
\Omega (\Sigma _1)-\Omega (\Sigma _2)=\int_VdV\delta \delta L-\delta
((E-L)_a\delta {\phi ^a})-\int_{\Sigma _3}d\Sigma _{3\mu }\delta j^\mu
\end{equation}
Obviously the first term vanishes because $\delta \delta =0$, the second
term vanishes because in solution space equations of motion $(E-L)_a$ are
satisfied, and the last term vanishes because the fields tend to zero fast
enough, so we have 
\begin{equation}
\Omega (\Sigma _1)-\Omega (\Sigma _2)=0,
\end{equation}
which ensure $\Omega $ is well defined on solution space.

Now we give the real scalar field as a simple examples to illustrate the
idea.

The Lagrangian density of scalar field theory is 
\begin{equation}
L= \frac 12(\partial ^\alpha \phi \partial _\alpha \phi -V(\phi )) ,
\label{scalar-Lag}
\end{equation}
from which the equation of motion 
\begin{equation}
\partial_{\mu}\partial^{\mu} \phi + V^{\prime }(\phi )=0.  \label{??}
\end{equation}
be obtained. Now take a solution of the equation of motion as a point of the
phase space $Z$. Thus we can define a function on $Z $ by $\phi (x):\phi
\longmapsto \phi (x)\in R$. The small displacements $\delta\phi $ is
determined by the linearized equation of motion. 
\begin{equation}
\partial_{\mu}\partial^{\mu} \delta \phi + V^{\prime \prime }(\phi )\delta
\phi =0.
\end{equation}
From eq. (\ref{X104})and(\ref{scalar-Lag}) the symplectic current is 
\begin{equation}
\delta J^\mu (x)=\delta (\frac{\partial L}{\partial (\partial _\mu \phi) }%
)\wedge \delta \phi (x)=\delta (\partial ^\mu \phi (x))\wedge \delta \phi
(x).  \label{  }
\end{equation}
Then the presymplectic form is 
\begin{equation}
\Omega =\int_{\Sigma }d\Sigma _\mu (x)\delta J^\mu (x)=\int_{\Sigma }d\Sigma
^\mu \delta (\partial _\mu \phi (x))\wedge \delta \phi (x),
\label{sym-sc--Malar}
\end{equation}
where $\Sigma $ is a space-like hypersurface of Minkowski space-time. Being
a closed and non-degenerate two form, the expression in (\ref{sym-sc--Malar}%
) gives a symplectic structure for scalar field on $Z $.

The above formulation has been used to the Yang-Mills theory\cite{Zuckman}%
\cite{CE} and general relativity and the gravitational WZW-model\cite{aldaya}
and the string theory\cite{Crnkovic2}. In the next section we will use it to
Palatini and Ashtekar gravity\cite{LSZ}.

\section{Geometric Quantization of Free Fields in  Space of Motions}
In the previous section, the classical symplectic description in  space of 
motions has
been established. In this section, we will complete geometric quantization
of free fields in solution space via K$\ddot{a}$hler polarization.

From the Lagrangian density of free real scalar field given in 
(\ref{scalar-Lag}), 
the Hamiltonian density 
\begin{equation}
h=\frac 12(\partial ^\alpha \phi \partial _\alpha \phi +m^2\phi ^2);~~~
\label{51s2}
\end{equation}
is obtained. 

The general solutions of equation of motion are: 
\begin{equation}
\varphi (x)=(2\pi )^{-3/2}\int \frac{d^3k}{\sqrt{2\epsilon (\vec{k})}}(a(%
\vec{k})e^{-ikx}+a^{*}(\vec{k})e^{ikx}),  \label{51s3}
\end{equation}
where $\epsilon (\vec{k})=k^0=\sqrt{\vec{k}^2+m^2}$ and $kx=k_\mu x^\mu
=k^0t-\vec{k}\cdot \vec{x}$ and $a(\vec{k})$, complex conjugate to $a^{*}(%
\vec{k})$, is 
\begin{equation}
\begin{array}{ll}
a(\vec{k}) & =\frac 1{\sqrt{2}}(k^0\varphi (\vec{k})+i\pi (\vec{k})) \\ 
& =(2\pi )^{-3/2}\int \frac{d^3x}{\sqrt{2\epsilon (\vec{k})}}[k^0\varphi
(x)+i\dot{\varphi}(x)]e^{ikx}.
\end{array}
\label{51s4}
\end{equation}
So the solution space $M_0$ can be coordinated by $(a^{*}(\vec{k}),a(\vec{k}%
))$. Putting the solution (\ref{51s3}) into (\ref{31presym}), we derive the
symplectic form in $M_0$ for real scalar fields 
\begin{equation}
\omega =i\int d^3k\delta a^{*}(\vec{k})\wedge \delta a(\vec{k}),
\label{51s5}
\end{equation}
where antisymmetry of the infinity exterior operator $\delta \wedge \delta $
have been used. Eq. (\ref{51s5}) explicitly shows that $\omega $ on
solutions space of free real scalar field is a well defined non-degenerate
closed 2-form . This expression is an analogue of the finite dimensional
harmonic oscillator whose symplectic form is 
\begin{equation}
\epsilon =dp_i\wedge dq^i=ida_i^{*}\wedge da_i.  \label{51s6}
\end{equation}
The Hamiltonian vector field of function $f$ is defined as 
\begin{equation}
i_{X_f}\omega =-\delta f,  \label{51sh}
\end{equation}
where $i$ here is used to represent the contraction between vector field and
differential forms in infinite space. So, the Hamiltonian vector field of $a(%
{\vec{k}})$ and $a^{*}({\vec{k}})$ can be obtained immediately 
\begin{equation}
X_{a(\vec{k})}=i\frac \delta {\delta a^{*}(\vec{k})},  \label{51s6a}
\end{equation}
\begin{equation}
X_{a^{*}(\vec{k})}=-i\frac \delta {\delta a(\vec{k})}.  \label{51s6b}
\end{equation}
(\ref{51s6a}) and (\ref{51s6b}) implies the basic Poisson bracket 
\begin{equation}
\{a^{*}(\vec{k}),a(\vec{k^{\prime }})\}=-i\omega (X_{a^{*}(\vec{k})},X_{a(%
\vec{k^{\prime }})})=X_{a^{*}(\vec{k})}a(\vec{k^{\prime }})=-i\delta ^3(\vec{%
k}-\vec{k^{\prime }}).
\end{equation}
which lead to commutator distribution\cite{BD} 
\begin{equation}
\begin{array}{l}
\bigtriangleup (x,y)=\{\varphi (x),\varphi (y)\}~~~~~~~~~~~~~~~~~~~~~ \\ 
=i(D(x-y)-D(y-x)) \\ 
=\frac 1{(2\pi )^3}\int \frac{d^3k(\vec{k})}{k^0}sink(x-y),\label{51dis}
\end{array}
\end{equation}
where 
\begin{equation}
D(x)=\frac 1{(2\pi )^3}\int \frac{d^3k(\vec{k})}{2k^0}e^{-ikx}.
\end{equation}

making use of solution of free equation (\ref{51s3}) and density expressions
(\ref{51s2}), the Hamiltonian become 
\begin{equation}
H=\int d^3k\epsilon (\vec{k})a^{*}(\vec{k})a(\vec{k}) , \label{51s8}
\end{equation}
whose Hamiltonian vector field is obtained as 
\begin{equation}
X_H=i\int d^3k\epsilon (\vec{k})[a^{*}(\vec{k})\frac \delta {\delta a^{*}(%
\vec{k})}-a(\vec{k})\frac \delta {\delta a(\vec{k})}].  \label{51s66}
\end{equation}

Next, we explicitly write down the symplectic structures of Maxwell field
with Lagrangian density 
\begin{equation}
L=\frac 14F_{\mu \nu }F^{\mu \nu }.
\end{equation}
with field strength $F_{\mu \nu }=\partial _\mu A_\nu -\partial _\nu A_\mu .$
The presymplectic form is then defined as 
\begin{equation}
\Omega =\int d\Sigma _\mu [\delta \partial ^\mu A^\nu \wedge \delta A_\nu
+\partial ^\nu \delta A^\mu \wedge \delta A_\nu ].  \label{presym-Maxwell0}
\end{equation}
Using the Lorentz condition $\partial _\mu A^\mu =0$ and, omitting the
integral on boundary, (\ref{presym-Maxwell0})  becomes 
\begin{equation}
\Omega =\int d\Sigma _\mu \delta \partial ^\mu A^\nu \wedge \delta A_\nu .
\label{presym-Maxwell}
\end{equation}
Make use of solutions of equations of motion 
\begin{equation}
A_\mu (x)=(2\pi )^{-3/2}\int \frac{d^3k}{\sqrt{2\epsilon (\vec{k})}}(c_\mu (%
\vec{k})e^{-ikx}+c_\mu ^{*}(\vec{k})e^{ikx}),  \label{52m1}
\end{equation}
the presymplectic form Eq. (\ref{presym-Maxwell}) can be written as 
\begin{equation}
\omega =i\int d^3k\delta c_\mu ^{*}(\vec{k})\wedge \delta c^\mu (\vec{k}).
\label{presym-Maxwell2}
\end{equation}
Similar to (\ref{51dis}) for scalar field, the commutator distribution are 
\begin{equation}
\{A_\mu (x),A_\nu (y)\}=i\eta _{\mu \nu }(D(x-y)-D(y-x)),  \label{52dis2}
\end{equation}
where $D(x-y)$ is defined as before.

As pointed out in the previous section, the presymplectic form (\ref
{presym-Maxwell}) degenerates along the $U(1)$ gauge orbits. Therefore only
when it restricts to module space, (\ref{presym-Maxwell}) can be defined as
symplectic form. In the solution (momentum) space Lorentz constraint becomes 
\begin{equation}
\begin{array}{ll}
k_\mu c^\mu (\vec{k})=0; & k_\mu c^{*\mu }(\vec{k})=0.\label%
{lorentz-constraint}
\end{array}
\end{equation}
We choose the polarized coordinate of photon 
\begin{equation}
k_\mu =(k,-k,0,0).  \label{pola-photon}
\end{equation}
Then we obtain from Eq. (\ref{pola-photon}) and (\ref{lorentz-constraint}) 
\begin{equation}
\begin{array}{ll}
c^0(\vec{k})=c^3(\vec{k})=-c_3(\vec{k}); & c^{*0}(\vec{k})=c^{*3}(\vec{k}%
)=-c_3^{*}(\vec{k}).
\end{array}
\end{equation}
So the symplectic form is 
\begin{equation}
\omega =i\int d^3k\delta c_i^{*}(\vec{k})\wedge \delta c_i(\vec{k}),~~~i=1,2,
\label{sym-Maxwell2}
\end{equation}
which implies only the transverse degrees of freedom have contribution to $%
\omega $ which is consistent with the degeneracy of $\omega $ along
non-physical components.

The third example is free Dirac field with Lagrangian density 
\begin{equation}
{\cal L}={\overline{\psi }}(i{\hat{\partial}}-m)\psi
\end{equation}
where ${\hat{\partial}}=\gamma ^\mu \partial _\mu $. Here $\psi $ and ${%
\overline{\psi }}$ are valued in Grassmannian numbers. The general solutions
of equation are 
\begin{eqnarray}
\psi (x) &=&\int \frac{d^3p}{(2\pi )^{3/2}}\frac 1{\sqrt{2E_p}}(e^{-ipx}a_i(%
\vec{p})u_i(\vec{p})+e^{ipx}b_i^{*}(\vec{p})v_i(\vec{p})  ,\label{52s49} \\
{\overline{\psi }(x)} &=&\int \frac{d^3p}{(2\pi )^{3/2}}\frac 1{\sqrt{2E_p}%
}(e^{ipx}a_i^{*}(\vec{p}){\overline{u}_i}(\vec{p})+e^{-ipx}b_i(\vec{p}){%
\overline{v}}_i(\vec{p}).  \label{52s50}
\end{eqnarray}
The solution manifold $M_F$ is Grassmannian\cite{kostant} which is
coordinated by $(a_i^{*},a_i,b_i^{*},b_i)$. The functional and vector field
and differential forms on Grassmannian manifold $M_F$ can also be defined.
For example, the Hamiltonian is 
\begin{equation}
H=\int d^3p\hbar E_p(a_i^{*}a_i+b_i^{*}b_i)
\end{equation}
The symplectic 2-form is then defined as 
\begin{equation}
\omega =\int d^3x\delta {\overline{\psi }}i\gamma ^0\delta \psi
 .
\label{52s51}
\end{equation}
substituting (\ref{52s49}) and (\ref{52s50}) in (\ref{52s51}) and using
relation 
\begin{eqnarray}
{\overline{u}}_i\gamma ^0u_j({\vec{p}}) &=&u_i^{+}u_j=2E_p\delta _{ij} \\
{\overline{v}}_i\gamma ^0u_j({\vec{p}}) &=&v_i^{+}v_j=2E_p\delta _{ij},
\end{eqnarray}
we obtain 
\begin{equation}
\omega =\int d^3p(\delta a_i^{*}\wedge \delta a_i+\delta b_i^{*}\wedge
\delta b_i)
\end{equation}
where we have used the antisymmetry of both differential forms and
Grassmanian numbers. The Hamiltonian vector field $X_f$ is defined as 
\begin{equation}
i_{X_f}\omega =-\delta f
\end{equation}
Then the anticommutator distribution can be obtained 
\begin{equation}
\{\psi (x),\psi (y)\}=i(i{\hat{\partial}}_x+m)(D(x-y)-D(y-x)).
\end{equation}

Now we perform the quantization of free real scalar field via K$\ddot a$hler
polarization. The positive polarization $P=F_B$ (B means Bargmann-Fock) is
spanned by the frame fields $X_{a^{}(\vec{k})}$ in Eq.(\ref{51s6a}). From
eqs(\ref{51s6a})(\ref{51s6b}) it is obvious that 
\begin{equation}
F_B\cap \overline{F_B}=0,~~F_B\cup \overline{F_B}={\cal T}^cX,
\end{equation}
where ${\cal T}^cX$ is the tangent space of the complexized phase space $X$
coordinated by $(k^0\varphi (\vec{k}),\pi (\vec{k}))$. So the polarization
is strongly admissible and complete. Being a manifold of $R^{6\infty }$, $X$
is contractible formally. The metalinear frame fields of $\overline{F_B}$ is
defined to be $\tilde{X}_{a^{*}(\vec{k})}$ who covers $X_{a^{*}(\vec{k})}$
and the half form bundle is defined to be $\delta _{-\frac 12}({\overline{P}}%
)$ whose section is denoted by $\nu _{\tilde{X}_{a^{*}(\vec{k})}}$
satisfying 
\begin{equation}
\nu _{\tilde{X}_{a^{*}(\vec{k})}}^{\#}\cdot \tilde{X}_{a^{*}(\vec{k})}=1,
\end{equation}
where $\#$ denotes the horizontal lift. Then we denote $(B_0,%
\bigtriangledown ,<,>)$ the prequantization structure with $B_0$ the trivial
line bundle and $\bigtriangledown $ the covariant derivative and $\lambda $
the trivialization section and $<,>$ the inner product in the fiber,
respectively. The prequantization structure satisfies 
\begin{equation}
\bigtriangledown \lambda _0=-i\frac 1\hbar \int d^3k\pi (\vec{k})\wedge
\delta (k^0\varphi (\vec{k}))\lambda _0,
\end{equation}
where $\lambda _0$ is unit section satisfying 
\begin{equation}
<\lambda _0,\lambda _0>=1.  \label{ilamda0}
\end{equation}
Introducing the new trivialization section $\lambda _1$ of $B_0$ by 
\begin{equation}
\lambda _1=exp[-\frac 1{4\hbar }\int d^3k((k^0\varphi (\vec{k}))^2+\pi (\vec{%
k})^2-2ik^0\varphi (\vec{k})\pi (\vec{k}))]\lambda _0,  \label{lamda1}
\end{equation}
we have 
\begin{equation}
\bigtriangledown \lambda _1=-i\frac 1\hbar \theta _1\lambda _1,
\label{cdlamda}
\end{equation}
where the connection $\theta _1$ is the symplectic potential by 
\begin{equation}
\theta _1=-i\int d^3ka(\vec{k})\delta a^{*}(\vec{k}).  \label{52c1}
\end{equation}
Eq.(\ref{cdlamda}) suggests that $\lambda _1$ is constant along ${\overline{F%
}_B}$, namely 
\begin{equation}
\bigtriangledown _{\overline{F_B}}\lambda _1=0.
\end{equation}
So, symplectic potential in (\ref{52c1}) is adapted to polarization $P$.
Clearly the section of $B_0\otimes \delta _{-\frac 12}({\overline{P}})$,
which is constant along ${\overline{F}_B}$, has the expression 
\begin{equation}
\sigma =\psi (a^{*}(\vec{k}))\lambda _1\otimes \nu _{\tilde{X}_{a^{*}(\vec{k}%
)}},
\end{equation}
where $\psi (a^{*}(\vec{k}))$ is the holomorphic functional of $a^{*}(\vec{k}%
)$. We define ${\cal H}$ the representation space corresponding to
polarization $F_B$ on which the scalar product is given by 
\begin{equation}
\begin{array}{l}
~~<\psi _1(a^{*}(\vec{k}))\lambda _1\otimes \nu _{\tilde{X}_{a^{*}(\vec{k}%
)}}|\psi _2(a^{*}(\vec{k}))\lambda _1\otimes \nu _{\tilde{X}_{a^{*}(\vec{k}%
)}}> \\ 
=\int_{R^{6\infty }}\psi _1^{*}(a^{*}(\vec{k}))\psi _2(a^{*}(\vec{k}%
))<\lambda _1,\lambda _1>\Pi \delta \varphi (\vec{k})\delta \pi (\vec{k}).
\end{array}
\label{scalarproduct}
\end{equation}
From Eqs. (\ref{ilamda0}) and(\ref{lamda1}) 
\begin{equation}
<\lambda _1,\lambda _1>=exp(-\frac 1\hbar \int d^3ka^{*}(\vec{k})a(\vec{k}))
\label{ilamda1}
\end{equation}
is obtained. Putting Eqs. (\ref{ilamda1}) into (\ref{scalarproduct}), one
gets 
\begin{equation}
\begin{array}{l}
~~~~<\psi _1(a^{*}(\vec{k}))\lambda _1\otimes \nu _{\tilde{X}_{a^{*}(\vec{k}%
)}}|\psi _2(a^{*}(\vec{k}))\lambda _1\otimes \nu _{\tilde{X}_{a^{*}(\vec{k}%
)}}> \\ 
=\int_{R^{6\infty }}\psi _1^{*}(a^{*}(\vec{k}))\psi _2(a^{*}(\vec{k}%
))exp(-\frac 1\hbar \int d^3ka^{*}(\vec{k})a(\vec{k}))\Pi \delta \varphi (%
\vec{k})\delta \pi (\vec{k}).
\end{array}
\end{equation}
This result can be also obtained by requiring $a^{*}(\vec{k})$ and $a(\vec{k}%
)$ to be self-adjoint on the Hilbert space.

In the Bargmann-Fock representation, like in the finite dimensional harmonic
oscillator, the observable functional $f$ has the representation 
\begin{equation}
\hat{O_f}=-i\hbar \bigtriangledown _{X_f}+f-i\hbar \frac 12\int d^3kn_0(\vec{%
k},\vec{k}),
\end{equation}
where $n_0(\vec{k},\vec{k})$ is determined by 
\begin{equation}
{\cal L}_{X_f}\delta a^{*}(\vec{k})=\int n_0(\vec{k},{\vec{k^{\prime }}}%
)\delta a^{*}({\vec{k^{\prime }}})d^3{\vec{k^{\prime }}}.
\end{equation}
Using (\ref{51s66}), one gets the ${\cal L}$i.e. derivative of $\delta a^{*}(%
\vec{k})$ along $X_H$ 
\begin{equation}
{\cal L}_{X_H}\delta a^{*}(\vec{k})=i\epsilon (\vec{k})\delta a^{*}(\vec{k}),
\end{equation}
which means $X_H$ preserve the polarization and gives 
\begin{equation}
n_0(\vec{k},{\vec{k^{\prime }}})=i\epsilon (\vec{k})\delta ^3(\vec{k}-{\vec{%
k^{\prime }}}),  \label{n0}
\end{equation}
or the ${\cal L}$i.e. derivative of the half form along $X_H$ is 
\begin{equation}
{\cal L}_{X_H}\sqrt{\delta a^{*}(\vec{k})}=\frac i2\epsilon (\vec{k})\sqrt{%
\delta a^{*}(\vec{k})}.
\end{equation}
So, in the polarization $P=F_B$, one has 
\begin{equation}
\hat{O_H}(\lambda _1\otimes \nu _{\tilde{X}_{a^{*}(\vec{k})}})=(-i\hbar
\bigtriangledown _{X_H}+H-i\hbar \frac 12\int d^3kn_0(\vec{k},\vec{k}%
))\lambda _1\otimes \nu _{\tilde{X}_{a^{*}(\vec{k})}}.  \label{O_H}
\end{equation}
From Eqs. (\ref{cdlamda})(\ref{51s66}) 
\begin{equation}
-i\hbar \bigtriangledown _{X_H}\lambda _1+H\lambda _1=0.  \label{X100}
\end{equation}
Then finally, using Eqs. (\ref{n0})(\ref{O_H}), one obtains 
\begin{equation}
\hat{O_H}(\psi (a^{*}(\vec{k}))\lambda _1\otimes \nu _{\tilde{X}_{a^{*}(\vec{%
k})}})=\int d^3k\epsilon (\vec{k})\hbar (N(\vec{k})+\frac 12\delta
^3(0))(\psi (a^{*}(\vec{k}))\lambda _1\otimes \nu _{\tilde{X}_{a^{*}(\vec{k}%
)}}).  \label{X101}
\end{equation}
in which $N(\vec{k})=a^{*}(\vec{k})\frac \delta {\delta a^{*}(\vec{k})}$ is
the operator representation of particles number. Due to the infinite
dimensions of degrees, the zero-point energy $\int d^3k\frac 12\epsilon (%
\vec{k})\delta ^3(0)$ is divergent, which can be removed by defining normal
product. Similarly, the observable momentum has the representation as
follows 
\begin{equation}
\hat{O}_{\vec{p}}=\int d^3k\vec{k}\hbar (N(\vec{k})+\frac 12\delta ^3(0)).
\end{equation}

In the same way, we can obtain the representation of field operator $%
O(\phi(x)$ and its quantum commutator 
\begin{equation}
[O_{(\phi(x)}, O_{(\phi(y)}]=\hbar( D(x-y)-D(y-x) )
\end{equation}
which is $(-i\hbar)$ times classical commutator distribution. We point out
here that since in solution space we use covariant symplectic structure
which preserve Poincare symmetry, and the procedure we take is in a
coordinate-free form, the quantum commutator must also preserve the Poincare
symmetry. In traditional canonical quantization, since the special time is
singled out, the Poincare invariance of commutators needs a proof\cite{PS}.
But here as we adopt the scheme preserving Poincare invariance from the very
beginning, the Poincare invariance of commutator is a natural result.

The procedures to deal with Maxwell field and free Dirac field can be
performed in a similar way as real scalar field, except considering
constraint in Maxwell field and properties of Grassmannian numbers in free
Dirac field. The results are the same as canonical quantization in
traditional phase space, which will not be listed here.

So far, the geometric quantization of free fields in solution space gives
the correct results as that obtained in traditional canonical quantization.
The remarkable property is that the Poincare covariance are kept all through
the procedure, just as path integral approach does.

\end{document}